\documentclass[nofigure]{pasj00}
\SetRunningHead{Distance to G192.16$-$3.84}{Shiozaki et al.}

\usepackage{times}
\usepackage[dvips]{graphicx,color}
\usepackage[utf8]{inputenc}

\newcommand{\myemail}{siozaki@milkiway.kagoshima-u.ac.jp}
\def\vlsr{$V_{\mbox{\scriptsize LSR}}$}
\def\kms{~km~s$^{-1}$}

\def\h2o{H$_{2}$O}

\def\g192{G192.16$-$3.84}

\begin{document}

\title{\h2o maser motions and the distance of the star forming region \g192}
\author{Satoshi  \textsc{Shiozaki}\altaffilmark{1}, Hiroshi  \textsc{Imai}\altaffilmark{1}, 
Daniel \textsc{Tafoya}\altaffilmark{1}, Toshihiro \textsc{Omodaka}\altaffilmark{1}, Tomoya \textsc{Hirota}\altaffilmark{2}, Mareki \textsc{Honma}\altaffilmark{2}, Makoto \textsc{Matsui}\altaffilmark{1}, and Yuji \textsc{Ueno}\altaffilmark{2}}

\altaffiltext{1}{Department of Physics and Astronomy, 
Graduate School of Science and Engineering, Kagoshima University, 1-21-35 Korimoto, Kagoshima 890-0065}

\altaffiltext{2}{Mizusawa VLBI Observatory, National Astronomical Observatory of Japan, 2-21-1 Osawa, Mitaka, Tokyo 181-8588}

\email{\myemail}
\KeyWords{ masers---stars: formation---stars: individual(\g192)}

\maketitle

\begin{abstract}
We present the results of  astrometic observations of \h2o masers associated with the star forming region \g192 with the VLBI Exploration of Radio Astrometry (VERA).
The \h2o masers seem to be associated with two young stellar objects (YSOs) separated by $\sim$1200 AU as reported in previous observations. In the present observations, we successfully detected an annual parallax of $\pi = 0.66\pm0.04$~mas for the \h2o masers, which corresponds to a distance to \g192 of $D = 1.52\pm 0.08$~kpc from the Sun. The determined distance is shorter than the estimated kinematic distance. Using the annual parallax distance and the estimated parameters of the millimeter continuum emission, we estimate the mass of the disk plus circumstellar cloud in the southern young stellar object to be 10.0$^{+4.3}_{-3.6}${\it M}$_{\odot}$. We also estimate the galactocentric distance and the peculiar motion of \g192, relative to a circular Galactic rotation: $R_{\ast}$ = 9.99 $\pm$ 0.08 kpc, $Z_{\ast} = -0.10 \pm 0.01$~kpc, and $(U_{\ast},  V_{\ast}, W_{\ast})= (-2.8 \pm 1.0, -10.5 \pm 0.3, 4.9 \pm 2.7)$ [km~s$^{-1}$] respectively. The peculiar motion of \g192 is within that typically found in recent VLBI astrometric results. The angular distribution and three-dimensional velocity field of \h2o\ maser features associated with the northern YSO indicate the existence of a bipolar outflow with a major axis along the northeast--southwest direction.  
\end{abstract}

\section{Introduction}

Measuring distances to astronomical objects with high accuracy allows us to unambiguously derive their physical parameters. 
Therefore, this is important especially for the legitimate identification of massive young stellar objects 
(MYSOs). Through a reliable estimation of the physical parameters of the MYSOs one enables to draw unambiguous conclusions 
about the most plausible scenarios for the formation of massive stars (see e.g., Cepheus  A, \cite{pat05,mos09}). To estimate the 
distances to star forming regions in the Galactic scale, the kinematic distance method has been commonly adopted. However, its 
accuracy significantly depends on how well we know the line-of-sight velocity of the source, as well as on the adopted 
Galactic constants and rotation curve (e.g., \cite{gom06}). In addition, ambiguity between the near and far distance emerges for the
observations towards the inner Galaxy. Furthermore, large deviations from a circular galactic rotation have been reported 
(e.g., \cite{xu06,rei09,bab09}). Thus, the estimated kinematic distances can be highly uncertain. The most straightforward method to 
determine the distance to astronomical objects is the trigonometric parallax method. Recent VLBI astrometric observations such as 
those carried out with, e.g., the Very Long Baseline Array (VLBA) and the VLBI Exploration of Radio Astrometry (VERA) have yielded 
trigonometric parallaxes on the kilo-parsec scale with errors less than 10\%. The latter array is mostly dedicated for astrometry of 
Galactic maser sources (e.g., \cite{hon07, hir07, nak08, oh10, sat10, ima11, nag11, nak11}). 

\begin{table*}[t]
\footnotesize
\begin{center}
\caption{Parameters of VERA observations  and data reduction.}
\label{tab:observations}
\vspace{2mm}
\begin{tabular}[b]{cccccccc} 
\hline
\hline 
  &
  &
  & Reference
  &
  & 
  & 
  & 
\\
  & Date
  & VERA
  & velocity$^{\ddagger}$
  & \multicolumn{2}{c}{Beam$^{\S}$} 
  & RMS noise
  &
\\
  Epoch$^{\ast}$
  &(yy/mm/dd)
  &telescopes$^{\dagger}$
  &(km s$^{-1}$)
  & \multicolumn{2}{c}{(mas,\ \ \ \arcdeg)} 
  &(Jy beam$^{-1}$)
  & N$_{f}$$^{\parallel}$
\\                                           
\hline
\dots&2006/10/21&MIZ$,$IRK$,$OGA$,$ISH&\dots&\dots&\ \ \dots&\dots&\dots\\
\dots&2006/11/16&MIZ$,$IRK$,$OGA$,$ISH&\dots&\dots&\ \ \dots&\dots&\dots\\
1&2006/12/12&MIZ$,$IRK$,$OGA$,$ISH&11.8&1.26$\times$0.73,&$-$41&0.1&5\\
2&2007/02/10&MIZ$,$IRK$,$OGA$,$ISH&11.8&2.24$\times$0.65,&$-$52&0.7&5\\
3&2007/03/23&MIZ$,$IRK$,$OGA$,$ISH&11.8&1.43$\times$0.79,&$-$45&0.4&9\\
4&2007/04/26&MIZ$,$IRK$,$OGA$,$ISH&11.8&1.91$\times$0.67,&$-$53&1.0&7\\
5&2007/05/22&MIZ$,$IRK$,$OGA$,$ISH&11.8&1.47$\times$0.76,&$-$45&0.5&8\\
6&2007/09/15&MIZ$,$IRK$,$OGA$,$ISH&11.8&1.99$\times$0.63,&$-$44&1.4&6\\
7&2007/12/23&MIZ$,$IRK$,$OGA$,$ISH&11.8&1.27$\times$0.92,&$-$41&0.2&5\\
8&2008/03/31&MIZ$,$IRK$,$OGA$,$ISH&$-$1.7\ \ &1.31$\times$0.90,&$-$43&0.1&11\\
\dots&2008/06/14&IRK$,$OGA$,$ISH&\dots&\dots&\ \ \dots&\dots&\dots\\
\dots&2008/06/21&MIZ$,$OGA$,$ISH&$-$1.7\ \ &1.18$\times$0.70,&$-$37&0.1&\dots\\
9&2008/07/07&MIZ$,$IRK$,$OGA$,$ISH&$-$1.7\ \ &1.27$\times$0.88,&$-$40&0.1&7\\
\hline
\end{tabular}
\end{center}
{
\hspace{2.7cm}\parbox[l]{12.5cm}{\footnotemark[${\ast}$]\hspace{0.5mm}\scriptsize Numbering the epoch when the astrometric result was adopted to estimate an annual 
parallax and maser relative proper motions.}\\
\hspace{2.7cm}\parbox[l]{12.5cm}{\footnotemark[${\dagger}$]\hspace{0.5mm}\scriptsize Telescopes participated in the observation. MIZ: VERA 20 m telescope at Mizusawa, Iwate, IRK: at Iriki, Kagoshima, OGA:~at Ogasawara Islands, Tokyo, and ISH: at Ishigakijima, Okinawa.}\\
\hspace{2.7cm}\parbox[l]{12.5cm}{\footnotemark[${\ddagger}$]\hspace{0.5mm}\scriptsize Local-standard-of-rest velocity of the spectral channel used for the phase reference in data calibration.}\\
\hspace{2.7cm}\parbox[l]{12.5cm}{\footnotemark[${\S}$]\hspace{0.5mm}\scriptsize Synthesized beam made with naturally weighted visibilities; major and minor axis lengths and position angle.}\\
\hspace{2.7cm}\parbox[l]{12.5cm}{\footnotemark[${\parallel}$]\hspace{0.5mm}\scriptsize Number of detected maser features.}
}
\end{table*}

\g192 (hereafter abbreviated as G192) is a widely studied object with a great potential to provide valuable insight in our understanding 
on the formation of intermediate and high mass stars.  Based on the kinematic distance ($\sim$2~kpc), it has been estimated that the ultracompact (UC) 
H{\rm II} region in G192 contains a young star with a luminosity of $L_{\ast}\sim 3\times 10^{3} L_{\odot}$ and a spectral type B2. 
The existence of a giant bipolar outflow with Herbig-Haro objects and a possible circumstellar disk has been reported (\cite{dev99}; \cite{she96}; 
\cite{she98}, hereafter SWSC98; \cite{she99}, hereafter SK99; \cite{she01}, hereafter SCK01; \cite{ind03}; \cite{she04}, hereafter 
SBCSK04). The \h2o masers are roughly grouped into two clusters separated by $\sim$0$\rlap{.}$\arcsec6 in the 
north--south direction (SK99; SBCSK04; \cite{ima06}, hereafter Paper I). The southern \h2o maser cluster exhibits an alignment perpendicular to the outflow. 
Based on \h2o maser and centimeter continuum observations, SK99 and SBCSK04 proposed that the southern \h2o maser features 
are associated with a flattened rotating gas torus (1000 AU) around a B2 star, in which the maser velocities are roughly consistent 
with those in a Keplerian disk. Observations with the Japanese VLBI Network (JVN) revealed relative bulk motions of the two maser 
features in the southern cluster. Paper I demonstrated that the \h2o maser features in the southern cluster 
are well explained by a model of an infalling-rotating disk with a radius of $\sim$1000 AU and a central stellar mass of 
5--10~$M_{\odot}$. Note that the size of the proposed rotating gas torus or disk is smaller than those detected so far around other 
early B stars (1700~AU around IRAS 20126$+$4104, \cite{cea99}; 1800 AU around AFGL~5142, \cite{zha02}). On the other hand, 
in the northern \h2o maser cluster, the distribution of maser features is aligned with the major axis of an outflow traced by the 3.6 cm 
continuum emission (SK99).

In this paper we report the results of astrometric observations of the G192 \h2o masers with VERA. Section 2 describes the 
VERA observation and the astrometric data analysis. Section 3 presents the obtained parallax distance and the secular motion of G192, yielded 
by tracing \h2o masers in the northern cluster. The angular distribution and the 3D motions of the individual 
maser features are also presented in this Section. Section 4 discusses the physical parameters of G192 and the secular motion 
of G192 in the Galaxy. 

\section{Observations and data reduction}

We have carried out VERA observations of the G192 \h2o masers at 13 epochs from December 2006 to July 2008. Table 
\ref{tab:observations} gives the summary of the observations. In the present study we have only used the data obtained in 9 epochs. 
In the rest of the observations, the weather conditions at some of the antennas sites were unfavorable. This resulted in much higher system noise temperatures at these antenna sites and the unsolved phase fluctuations in these observations degraded the image qualities of G192 in phase reference analyses. For each epoch, 
the observation duration was  8--10 hours, including scans on G192 and the calibrators (J053056.4$+$133155, 3C 84 and DA 193). 
We observed the extragalactic quasar J060309.1$+$174216 (hereafter J0603) simultaneously with G192 using VERA's dual-beam 
system for the phase referencing (e.g., \cite{hon03, hon08}). The angular separation between G192 and J0603 is 1\arcdeg.66. 
The flux density of J0603 was $\sim$100 mJy  with a small variation ($<$20\%). The coordinates of G192 and J0603 used as 
reference for the observation and the data correlation were set to ($\alpha_{2000}$, $\delta_{2000}$) $=$ 
(05$^h$58$^m$13.$^s$53, $+$16$^\circ$31$'$58.$''$9) and ($\alpha_{2000}$, $\delta_{2000}$)  = 
(06$^{h}$03$^{m}$09$.^{s}$130269, $+$17$^\circ$42$'$16.$''$81070), respectively. 

\begin{figure*}[t!]
\begin{center}
\includegraphics[angle=-0,scale=0.8]{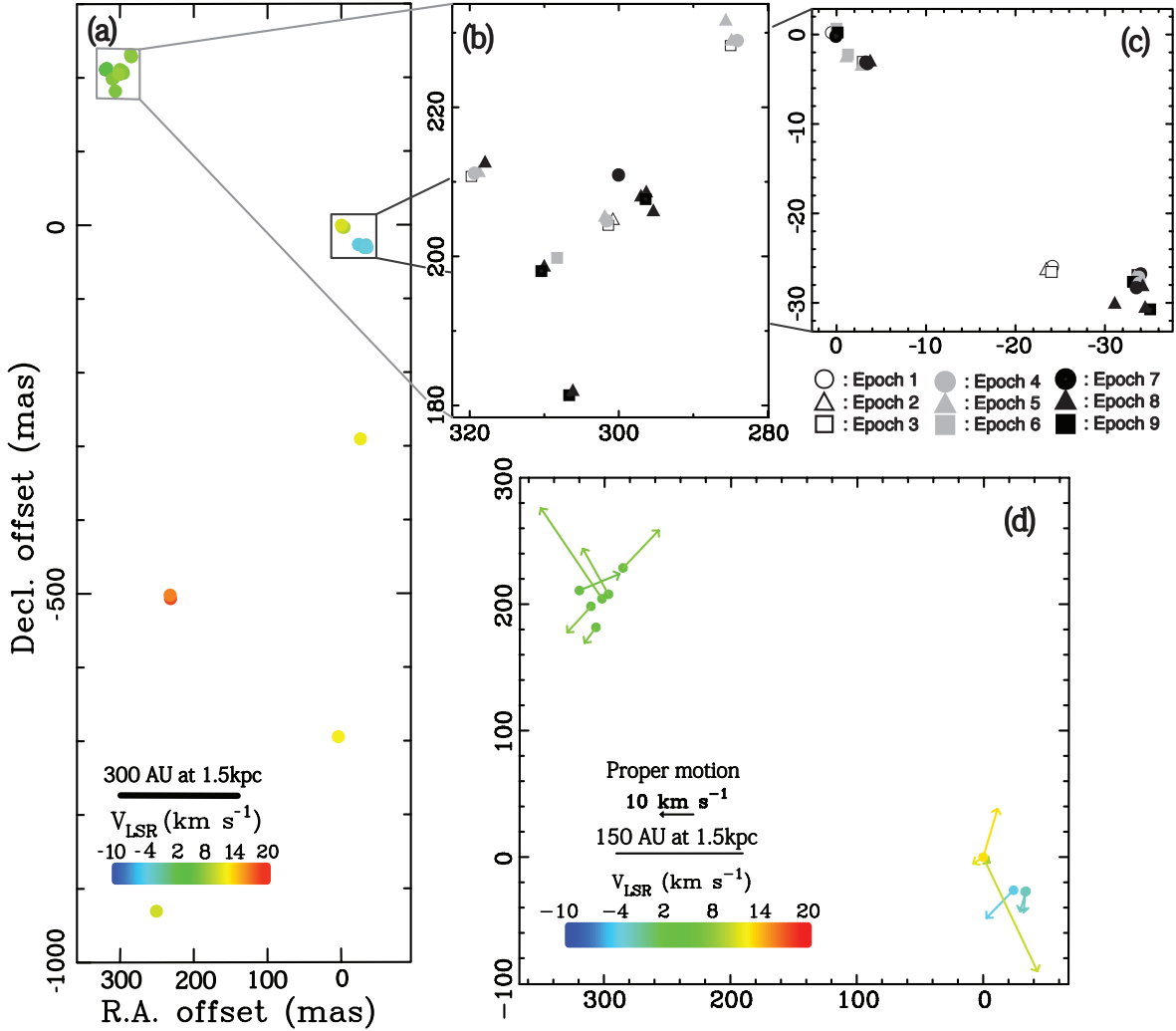}
\end{center}
\caption{
Distribution of \h2o maser features in  \g192. The coordinate origin is located at the position of Feature A at Epoch 1. Colors of 
maser features indicate LSR velocities. (a) Maser feature distribution in the whole observed area. The bold black arrows indicate the 
average proper motions of the NE and SW clusters of the \h2o maser features in the northern region. (b) Zoom-up view of the NE 
maser feature cluster. The different symbols denote the different observation epochs, whose ID number is identical with that listed 
in Table \ref{tab:proper-motions}. (c) Same as (b) but towards the SW maser feature cluster. (d)Relative proper motions of \h2o 
maser features (colored arrows) in  \g192 detected during the 9 epochs. The root of an arrow indicates the location of the feature 
at the first of the epochs when the maser feature was detected. The length and the direction of an arrow indicate the speed and 
direction of the maser proper motion, respectively. The bold arrows {\boldmath $V_{\rm NE}$} and {\boldmath $V_{\rm SW}$} are 
the same as the bold black arrows of (a).
}
\label{fig:1}
\end{figure*}

The received left-hand circular polarization signals were split into 16 base band channels with a band width of 16 MHz each and recorded at a rate of 1024 Mbit s$^{-1}$. One of the base band channels was assigned to the observations of the maser source while the other fifteen were assigned to the 
reference source. The data correlation was carried out with the Mitaka FX correlator. The accumulation period of the correlation 
was set to 1~s. The correlation outputs consisted of 512 spectral channels, yielding a velocity spacing of 0.42\kms. 

Most of the data reduction was carried out using the National Radio Astronomy Observatory's Astronomical Image Processing 
System (AIPS) package. In order to obtain, the final maser image cubes with sensitivity and quality higher than those obtained 
in the phase-referencing technique (see below), we adopted the {\it standard procedure} (e.g., \cite{dia95}). This procedure consists 
of performing the fringe-fitting and self-calibration on the data of a Doppler-velocity channel that includes the brightest \h2o maser 
spot (velocity channel component) in G192, which is used as a fringe-phase reference. Column 4 of Table \ref{tab:observations} 
gives the line-of-sight velocity of the reference spectral channel with respect to the Local Standard of Rest (LSR) in each epoch. 
Thus, all maser spot positions were measured with respect to that of the reference spot. We found the location of each maser spot 
using the AIPS task SAD, which fits a two-dimensional Gaussian model to the brightness distribution in the map. The 
detection limit of a maser spot was typically 500~mJy beam$^{-1}$ at a 7-$\sigma$ noise level when no other bright maser was present 
in the same velocity channel map. The position of each \h2o maser feature (a physical maser clump consisting of a cluster of maser spots or velocity components in a
physical gas clump) was defined as the position calculated from an intensity-weighted mean of maser spot positions in the feature. We found that the typical spread of maser spot positions in each maser feature was $\sim$0.2 mas , which may give an upper limit to the wander of feature position determined by an intensity-weighted mean of spot positions. Column 7 of Table \ref{tab:observations} gives the number of identified maser 
features in each epoch. Table \ref{tab:features} gives the list of all maser features detected in the observations.

In order to perform the maser source astrometry, some special procedures are needed (e.g., \cite{hir07, ima11}). 
Firstly, the delay-tracking was reperformed for the correlated data by using better delay-tracking models obtained using software 
equivalent to the CALC9 package developed by the Goddard Space Flight Center/NASA VLBI group. Secondly, after the visibility 
amplitude calibration was made, independently for both the maser and reference source data, we used the scans on the phase-reference 
source J0603 to obtain the group delay and fringe-phase residuals of the fringe fitting using the AIPS task FRING. The solution 
interval was usually set to 1~min, sometimes to 2 or 3~min according to the number of available antennas and the system temperatures.  
Thirdly, in order to calibrate the visibility data of G192, we copied the calibration solutions table from the J0603 data into those of 
G192 using the AIPS task TACOP. Also, fringe phase and amplitude solutions obtained by self-calibration of the J0603 data were 
applied to the G192 data.  As a result, we obtained the absolute positions of the bright spots at \vlsr=$-$1.7 -- 12.6 km/s with respect to the reference source. We regarded detection of a maser spot as true when the spot was detected in more than 4 epochs. Note that no special correction for source parallactic angle rotation has been performed. 
The VERA's receiving system is rotated so as to fix the 
fields of view with respect to the parallactic angles. 
In this case, it is expected that the difference in the linear polarization characteristics towards target 
and reference sources should be constant during a single observation session and the whole 
monitoring observations. 

\begin{center}
\footnotesize{
\begin{longtable}[t!]{r@{\hspace{0.8cm}}r@{$\pm$}lr@{$\pm$}l@{\hspace{-0.3cm}}rrc}

\caption{Parameters of the detected H$_{2}$O maser features.} \label{tab:features} \\
\hline\hline 
\multicolumn{8}{c}{\phantom{.}\hspace{6cm}\phantom{.}}\\[-2ex]
\multicolumn{1}{c}{v$_{\rm LSR}$$^{\ast}$} & 
\multicolumn{4}{c}{Offset$^{\dagger}$(mas)} & 
\multicolumn{1}{c}{$I^{\ddagger}$ } & 
\multicolumn{1}{c}{$\Delta$$V^{\S}$} &
\multicolumn{1}{c}{Feature}\\
\multicolumn{1}{c}{(km s$^{-1}$)} & 
\multicolumn{4}{c}{\ \hrulefill \ } & 
\multicolumn{1}{c}{(Jy beam$^{-1}$)} & 
\multicolumn{1}{c}{(km s$^{-1}$)} &
\multicolumn{1}{c}{ID$^{\parallel}$}\\
\multicolumn{1}{c}{} & 
\multicolumn{2}{c}{R.A.} & 
\multicolumn{2}{c}{Dec.} & 
\multicolumn{1}{c}{} &
\multicolumn{1}{c}{} &
\multicolumn{1}{c}{}\\
\hline
\endfirsthead

\hline\hline 
\multicolumn{8}{c}{\phantom{.}\hspace{6cm}\phantom{.}}\\[-2ex]
\multicolumn{1}{c}{v$_{\rm LSR}$$^{\ast}$} & 
\multicolumn{4}{c}{Offset$^{\dagger}$(mas)} & 
\multicolumn{1}{c}{$I^{\ddagger}$ } & 
\multicolumn{1}{c}{$\Delta$$V^{\S}$} &
\multicolumn{1}{c}{Feature}\\
\multicolumn{1}{c}{(km s$^{-1}$)} & 
\multicolumn{4}{c}{\ \hrulefill \ } & 
\multicolumn{1}{c}{(Jy beam$^{-1}$)} & 
\multicolumn{1}{c}{(km s$^{-1}$)} &
\multicolumn{1}{c}{ID$^{\parallel}$}\\
\multicolumn{1}{c}{} & 
\multicolumn{2}{c}{R.A.} & 
\multicolumn{2}{c}{Dec.} & 
\multicolumn{1}{c}{} &
\multicolumn{1}{c}{} &
\multicolumn{1}{c}{}\\
\hline 
\endhead

\multicolumn{8}{c}{}\\[-2ex]
\endfoot

\hline 

\multicolumn{8}{c}{}\\[-2ex]
\multicolumn{8}{p{10cm}}{\parbox[l]{12.4cm}{\footnotemark[${\ast}$ ]\hspace{0.5mm}\scriptsize Local-standard-rest velocity at the intensity peak.}} \\
\multicolumn{8}{p{10cm}}{\parbox[l]{12.4cm}{\footnotemark[${\dagger}$]\hspace{0.5mm} \scriptsize Position offset with respect to the brightest \h2o maser spot whose Doppler-velocity channel was used for fringe fitting and self-calibration.}} \\
\multicolumn{8}{p{10cm}}{\parbox[l]{12.4cm}{\footnotemark[${\ddagger}$]\hspace{0.5mm} \scriptsize Peak intensity of the maser feature.}}\\
\multicolumn{8}{p{10cm}}{\parbox[l]{12.4cm}{\footnotemark[${\S}$]\hspace{0.5mm} \scriptsize Full  velocity width in which the maser emission was detected. The minimum value equals to the velocity spacing of a spectral channel (0.42~km~s$^{-1}$).}}\\
\multicolumn{8}{p{4cm}}{\parbox[l]{12.4cm}{\footnotemark[${\parallel}$]\hspace{0.5mm} \scriptsize Ordinal number of the identification label for the maser features listed in Table \ref{tab:proper-motions}. The ID is designated as  G192:S2011-$n$, where S2011 indicates that we are referring to the maser features 
found in the present work, and $n$ is the ordinal source number.}}

\endlastfoot

%
 \multicolumn{8}{c}{2006 December 12} \\                                                 
\hline 
$   19.00$&$   231.650$& 0.050 &$  -506.480$& 0.060 &        1.46 &   0.42 & \dots\\
$   11.90$&$     0.042$& 0.204 &$    -0.056$& 0.053 &        7.68 &   2.53 &12\\
$   12.60$&$     3.980$& 0.070 &$  -693.700$& 0.050 &        0.77 &   0.42 &\dots\\
$   -1.70$&$   -33.840$& 0.040 &$   -26.990$& 0.040 &        2.19 &   0.42 &3\\
$   -3.40$&$   -23.810$& 0.041 &$   -26.260$& 0.041 &        1.76 &   0.42 &1\\
\hline
  \multicolumn{8}{c}{2007 February 10} \\                                                 
\hline 
$   11.82$&$     0.021$& 0.134 &$    -0.012$& 0.022 &       22.21 &   2.95 &12\\
$   12.22$&$   -26.020$& 0.220 &$  -289.790$& 0.100 &        4.85 &   0.42 &\dots\\
$    9.54$&$    -3.118$& 0.056 &$    -3.293$& 0.059 &        1.37 &   0.84 &10\\
$    8.85$&$   301.130$& 0.090 &$   204.590$& 0.070 &        0.91 &   0.42 &\dots\\
$   -3.37$&$   -23.580$& 0.050 &$   -26.420$& 0.040 &        1.40 &   0.42 &1\\
\hline
  \multicolumn{8}{c}{2007 March 23} \\                                                 
\hline 
$   13.49$&$     0.100$& 0.050 &$    -0.170$& 0.040 &        0.66 &   0.42 &13\\
$   10.84$&$     0.157$& 0.017 &$    -0.037$& 0.012 &        6.94 &   2.53 &11\\
$   11.71$&$     0.011$& 0.040 &$    -0.003$& 0.011 &       36.89 &   2.11 &12\\
$    9.68$&$    -2.855$& 0.039 &$    -3.367$& 0.016 &        2.51 &   1.26 &10\\
$    7.86$&$   301.397$& 0.020 &$   204.271$& 0.020 &        1.18 &   2.11 &7\\
$    8.01$&$   284.730$& 0.040 &$   228.720$& 0.040 &        1.00 &   0.42 &8\\
$    4.36$&$   319.346$& 0.129 &$   210.851$&0.093 &        0.80 &   0.84 &4\\
$   -1.68$&$   -33.590$& 0.051 &$   -27.210$& 0.070 &        0.56 &   0.42 &3\\
$   -3.25$&$   -23.790$& 0.087 &$   -26.411$& 0.070 &        0.70 &   0.84 &1\\
\hline
 \multicolumn{8}{c}{2007 April 26}\\                                                 
\hline 
$   13.49$&$     0.000$& 0.080 &$     0.050$& 0.080 &        1.20 &   0.42 &13\\
$   11.51$&$     0.009$& 0.026 &$    -0.005$& 0.017 &       42.06 &   2.95 &12\\
$    9.67$&$    -2.926$& 0.045 &$    -3.355$& 0.045 &        8.31 &   1.26 &10\\
$    8.77$&$   301.607$& 0.010 &$   204.573$& 0.063 &        5.93 &   1.27 &7\\
$    8.01$&$   284.440$& 0.050 &$   229.010$& 0.080 &        3.07 &   0.42 &8\\
$    4.37$&$   319.263$& 0.040 &$   210.955$& 0.040 &        2.83 &   0.84 &4\\
$   -2.05$&$   -33.492$& 0.152 &$   -27.383$& 0.072 &        1.97 &   1.26 &3\\
\hline
  \multicolumn{8}{c}{2007 May 22}\\                                                 
\hline 
$   11.49$&$     0.021$& 0.045 &$     0.002$& 0.020 &       44.04 &   2.53 &\dots\\
$   11.80$&$    -1.140$& 0.130 &$    -2.500$& 0.100 &        1.25 &   0.42 &12\\
$    9.68$&$    -2.968$& 0.043 &$    -3.345$& 0.045 &       13.82 &   1.26 &10\\
$    8.85$&$   301.720$& 0.090 &$   205.100$& 0.102 &        1.15 &   0.42 &7\\
$    7.89$&$   285.677$& 0.093 &$   231.465$& 0.030 &        3.85 &   1.26 &8\\
$    8.01$&$   284.570$& 0.100 &$   228.930$& 0.118 &        1.74 &   0.42 &\dots\\
$    4.26$&$   319.087$& 0.030 &$   211.031$& 0.030 &        2.25 &   1.27 &4\\
$   -2.11$&$   -33.170$& 0.090 &$   -27.610$& 0.091 &        1.04 &   0.42 &2\\
\hline
  \multicolumn{8}{c}{2007 September 15} \\                                                 
\hline 
$   11.24$&$     0.003$& 0.026 &$     0.002$& 0.063 &       27.29 &   1.69 &12\\
$   10.96$&$    -1.200$& 0.050 &$    -2.300$& 0.060 &        3.64 &   0.42 &11\\
$   10.96$&$   250.530$& 0.050 &$  -930.390$& 0.050 &        5.08 &   0.42 &\dots\\
$    9.73$&$    -3.175$& 0.020 &$    -3.280$& 0.066 &       10.27 &   1.26 &10\\
$   10.11$&$     0.020$& 0.040 &$     0.650$& 0.040 &        3.93 &   0.42 &\dots\\
$    8.01$&$   308.270$& 0.060 &$   199.800$& 0.090 &        2.00 &   0.42 &\dots\\
\hline
  \multicolumn{8}{c}{2007 December 23} \\                                                 
\hline 
$   11.34$&$    -0.001$& 0.069 &$    -0.012$& 0.190 &        9.16 &   2.95 &12\\
$    9.58$&$    -3.342$& 0.025 &$    -3.301$& 0.107 &        3.71 &   1.69 &10\\
$    7.73$&$   296.351$& 0.020 &$   207.804$& 0.066 &        1.86 &   0.84 &6\\
$    8.01$&$   300.030$& 0.040 &$   210.960$& 0.050 &        1.03 &   0.42 &\dots\\
$   -1.83$&$   -33.705$& 0.057 &$   -28.041$& 0.010 &        5.47 &   1.27 &3\\
\hline 
 \multicolumn{8}{c}{2008 March 31} \\                                                 
\hline 
$   11.49$&$    33.8210$& 0.027 &$    27.9480$& 0.095 &        4.55 &   2.11 &12\\
$    9.58$&$    30.156$& 0.117 &$    24.887$& 0.074 &        1.67 &   1.26 &10\\
$    7.47$&$   340.158$& 0.125 &$   209.505$& 0.010 &        4.40 &   1.26 &5\\
$    8.01$&$   343.930$& 0.080 &$   226.320$& 0.090 &        0.85 &   0.42 &9\\
$    7.59$&$   329.180$& 0.060 &$   233.950$& 0.102 &        1.14 &   0.42 &\dots\\
$    7.40$&$   330.359$& 0.050 &$   236.376$& 0.131 &        1.06 &   0.84 &6\\
$    7.59$&$   330.450$& 0.050 &$   235.930$& 0.074 &        0.85 &   0.42 &\dots\\
$   -1.84$&$    -0.004$& 0.023 &$    -0.000$& 0.001 &       97.51 &   2.53 &3\\
$    4.86$&$   351.742$& 0.129 &$   240.353$& 0.030 &        1.40 &   0.84 &\dots\\
$   -0.84$&$    -0.890$& 0.070 &$    -2.680$& 0.060 &        0.61 &   0.42 &\dots\\
$   -2.11$&$     2.630$& 0.040 &$    -2.230$& 0.080 &        2.63 &   0.42 &\dots\\
\hline
 \multicolumn{8}{c}{2008 July 07} \\                                                 
\hline 
$   16.92$&$   265.785$& 0.020 &$  -473.482$& 0.020 &        2.89 &   1.26 &\dots\\
$   11.33$&$    34.087$& 0.102 &$    28.148$& 0.120 &        2.19 &   1.26 &12\\
$    7.59$&$   340.440$& 0.040 &$   209.600$& 0.070 &        1.18 &   0.42 &5\\
$    7.59$&$   344.360$& 0.102 &$   226.290$& 0.050 &        1.44 &   0.42 &9\\
$   -1.75$&$     0.008$& 0.024 &$    -0.003$& 0.005 &       62.44 &   1.26 &3\\
$   -2.53$&$     0.490$& 0.040 &$    -0.140$& 0.060 &        1.33 &   0.42 &2\\
$   -2.95$&$    -0.940$& 0.050 &$    -2.570$& 0.050 &        1.35 &   0.42 &\dots\\

\end{longtable}

}

\end{center}

\section{Results}

\subsection{Relative proper motions of the \h2o\ masers}

Panel (a) of Figure \ref{fig:1} shows the angular distribution of the \h2o\ maser features in G192 found in this paper. They were located
in an area of 0$\rlap{.}$\arcsec7$\times$1$\rlap{.}$\arcsec3 and their spatial distribution of roughly reproduced those 
found in the previous VLA/VLBA/JVN observations (SK99; SBCSK04; Paper I). Towards the northern region, the maser 
features appear to be grouped in two clusters. The north-eastern cluster is redshifted with respect to the cluster located in the 
south-west direction. Panels (b) and (c) show the time variation of the angular distribution of the maser features in these clusters. Some 
of the maser features were detected in the southern region. Their velocity distribution is consistent with that proposed in the 
infalling-rotating disk model in Paper I. However, it is difficult to discuss the plausible velocity field in more detail in this paper due to a small 
number of the detected maser features and the lack of information of their relative proper motions. 

For the northern clusters, we detected relative proper motions of twelve \h2o maser features with respect to a reference feature 
labeled as G192:S2011-{\it12} (see the footnote in Table \ref{tab:features} for the explanation of the nomenclature). 
The proper motions were measured from those maser features that were detected in at least two epochs and that showed a 
displacement of $\lesssim$10 mas within a velocity drifts of $\lesssim$1\kms. Table \ref{tab:proper-motions} gives the parameters of the proper 
motions of twelve maser features. To estimate the maser proper motions with respect to the 
northern YSO in G192 (associated with the norther maser clusters), we assume that from the reference in which the YSO is fixed the 
vectorial sum of the proper 
motions of all the maser features equals zero. In order to do this, first we obtained the average of the proper motions of the masers 
with respect to the reference feature: ($\bar{\mu}_{x}$,~$\bar{\mu}_{y}$)=($-$0.36$\pm$0.14, 0.30$\pm$0.14) [mas~yr$^{-1}$]. Then, 
by subtracting it from (${\mu}_{x}$,~${\mu}_{y}$), we obtained the proper motions with respect to the reference frame that will move together with the G192 system. Panel 
(d) of Figure~\ref{fig:1} shows the three-dimensional velocity field of the maser features with respect to the northern YSO. The 
average expansion velocity of the masers in the jet is around v$_{\rm exp}$$\sim$17\kms\ (considering a distance of $D = 1.52$~kpc, 
see below), although there are a few of maser features that move as fast as $\sim$40\kms. As mentioned above, the proper motions 
of the southern maser features found in previous observations could not be reconfirmed by the present observations most probably 
due to their very short lifetimes.

\begin{table*}[t]
\footnotesize
\begin{center}
\caption{Relative proper motions of H$_{2}$O masers in  \g192.}
\label{tab:proper-motions}
\vspace{2mm}
\begin{tabular}[b]{lrrrrrrrr   c@{ }c@{ }c@{ }c@{ }c@{ }c@{ }c@{ }c@{ }c } \hline \hline
Maser feature$^{\ast}$
 & \multicolumn{2}{c}{Offset$^{\dagger}$}
 & \multicolumn{4}{c}{Proper motion$^{\dagger}$}
 & \multicolumn{2}{c}{Radial motion$^{\ddagger}$}
 & \\   
 (G192:S2011-$n$)                                                                                          
 & \multicolumn{2}{c}{(mas)} 
 & \multicolumn{4}{c}{(mas yr$^{-1}$)}
 & \multicolumn{2}{c}{(km s$^{-1}$)}
 & \multicolumn{ 9}{c}{Detection for 9 epochs$^{\S}$} \\                                                
 & \multicolumn{2}{c}{\ \hrulefill \ } 
 & \multicolumn{4}{c}{\ \hrulefill \ } 
 & \multicolumn{2}{c}{\ \hrulefill \ } 
 & \multicolumn{ 9}{c}{\ \hrulefill \ } \\                                                     
 $n$& R.A. & decl. & $\mu_{x}$ & $\sigma \mu_{x}$ & $\mu_{y}$ 
 & $\sigma \mu_{y}$ & $V_{z}$ & $\Delta$$V_{z}$$^{\parallel}$
 &1& 2& 3& 4& 5& 6& 7& 8& 9 \\ \hline
 1\ \dotfill \  &$   -23.85$&$   -26.20$&$   0.75$&   0.29 &$  -0.93$&   0.26
 &$  -3.37$&   0.42
 &$\circ$&$\circ$&$\circ$&$\times$&$\times$&$\times$&$\times$&$\times$&$\times$\\         
 2\ \dotfill \  &$   -33.19$&$   -27.61$&$  -0.36$&   0.09 &$  -0.60$&   0.10
 &$  -2.11$&   0.21
 &$\times$&$\times$&$\times$&$\times$&$\circ$&$\times$&$\times$&$\times$&$\circ$\\        
 3\ \dotfill \  &$   -33.88$&$   -26.93$&$  -0.16$&   0.03 &$  -0.60$&   0.01
 &$  -1.68$&   1.12
 &$\circ$&$\times$&$\circ$&$\circ$&$\times$&$\times$&$\circ$&$\circ$&$\circ$\\            
 4\ \dotfill \  &$   319.34$&$   210.85$&$  -2.17$&   0.55 &$   1.00$&   0.48
 &$   4.36$&   0.98
 &$\times$&$\times$&$\circ$&$\circ$&$\circ$&$\times$&$\times$&$\times$&$\times$\\         
 5\ \dotfill \  &$   306.34$&$   181.56$&$   0.06$&   0.49 &$  -0.39$&   0.26
 &$   7.47$&   0.74
 &$\times$&$\times$&$\times$&$\times$&$\times$&$\times$&$\times$&$\circ$&$\circ$\\        
 6\ \dotfill \  &$   296.35$&$   207.82$&$   0.69$&   0.20 &$   2.26$&   0.54
 &$   7.73$&   0.84
 &$\times$&$\times$&$\times$&$\times$&$\times$&$\times$&$\circ$&$\circ$&$\times$\\        
 7\ \dotfill \  &$   301.39$&$   204.27$&$   2.24$&   0.23 &$   4.23$&   0.48
 &$   7.86$&   1.20
 &$\times$&$\times$&$\circ$&$\circ$&$\circ$&$\times$&$\times$&$\times$&$\times$\\         
 8\ \dotfill \  &$   284.72$&$   228.72$&$  -1.97$&   0.52 &$   1.91$&   0.63
 &$   8.01$&   0.21
 &$\times$&$\times$&$\circ$&$\circ$&$\circ$&$\times$&$\times$&$\times$&$\times$\\         
 9\ \dotfill \  &$   310.11$&$   198.37$&$   0.61$&   0.48 &$  -0.85$&   0.38
 &$   8.01$&   0.21
 &$\times$&$\times$&$\times$&$\times$&$\times$&$\times$&$\times$&$\circ$&$\circ$\\        
 10\ \dotfill \  &$    -3.14$&$    -3.28$&$  -0.52$&   0.04 &$   0.22$&   0.06
 &$   9.54$&   1.26
 &$\times$&$\circ$&$\circ$&$\circ$&$\circ$&$\circ$&$\circ$&$\circ$&$\times$\\             
 11\ \dotfill \  &$     0.15$&$    -0.03$&$  -2.79$&   0.11 &$  -4.69$&   0.13
 &$  10.84$&   1.37
 &$\times$&$\times$&$\circ$&$\times$&$\times$&$\circ$&$\times$&$\times$&$\times$\\        
 12$^{\sharp}$\ \dotfill \  &$     0.00$&$     0.00$&$   0.00$&   0.03 &$   0.00$&   0.05
 &$  11.86$&   2.34
 &$\circ$&$\circ$&$\circ$&$\circ$&$\circ$&$\circ$&$\circ$&$\circ$&$\circ$\\               
 13\ \dotfill \  &$     0.09$&$    -0.17$&$  -1.06$&   1.02 &$   2.39$&   0.96
 &$  13.49$&   0.21
 &$\times$&$\times$&$\circ$&$\circ$&$\times$&$\times$&$\times$&$\times$&$\times$\\        
\hline
\end{tabular}
\end{center}
{
\hspace{1.8cm}\parbox[l]{14.5cm}{\footnotemark[${\ast}$]\hspace{0.5mm}\scriptsize H$_{2}$O maser features detected toward  \g192. The feature ID is designated as  G192:S2011-$n$, where S2011 indicates that 
we are referring to the maser features found in this work, and $n$ is the ordinal source ID number given in this column.}\\
\hspace{1.8cm}\parbox[l]{14.5cm}{\footnotemark[${\dagger}$]\hspace{0.5mm}\scriptsize Relative value with respect to the motion of the position-reference maser feature$:$  
\g192: S2011 {\it1}. }\\
\hspace{1.8cm}\parbox[l]{14.5cm}{\footnotemark[${\ddagger}$]\hspace{0.5mm}\scriptsize Relative value with respect to the local standard of rest. }\\
\hspace{1.8cm}\parbox[l]{14.5cm}{\footnotemark[${\S}$]\hspace{0.5mm}\scriptsize $\circ$$:$detection, $\times$$:$negative detection. }\\
\hspace{1.8cm}\parbox[l]{14.5cm}{\footnotemark[${\parallel}$]\hspace{0.5mm}\scriptsize Mean full velocity width of a maser feature at half intensity. }\\
\hspace{1.8cm}\parbox[l]{14.5cm}{\footnotemark[${\natural}$]\hspace{0.5mm}\scriptsize Reference feature for the relative proper motion measurement.}\\
}
\end{table*}

\begin{figure*}[t!]
\begin{center}
\includegraphics[angle=-0,scale=0.6]{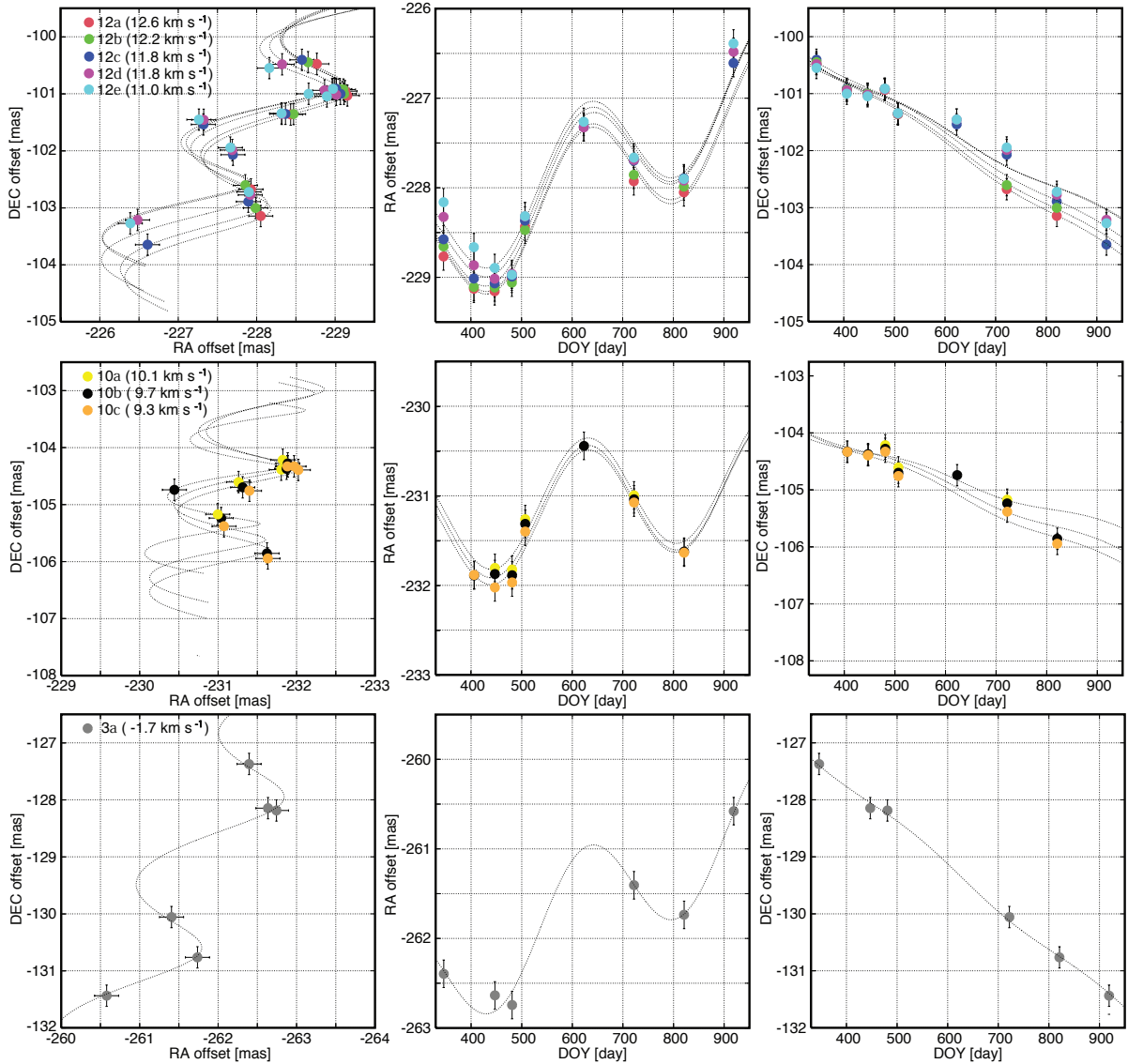}
\end{center}
\caption{
Motions of the \h2o\ maser spots and the best-fit results. The spot IDs correspond to those in Table \ref{tab:proper-motions}. 
(Left): the maser positions with respect to the phase-tracking center in the (R.A, Decl.) coordinates changing with time. A solid 
curve indicates the spot motion expected from the best-fit. (Middle and right): R.A. and Decl. offsets against time, respectively. 
}
\label{fig:2}
\end{figure*}

\begin{table*}[t!]
\footnotesize
\begin{center}
\caption
{\small Results of model fits of an annual parallax and a linear proper motion to the maser spots in  \g192.}
\label{tab:astrometry}
\vspace{2mm}
\begin{tabular}[c]{lr c@{ }c@{ }c@{ }c@{ }c@{ }c@{ }c@{ }c@{ }c ccllll} \hline \hline
 Spot ID$^{\ast}$
 &V$_{LSR}$ \ \ 
 & \multicolumn{9}{c}{Detection at 9 epochs$^{\S}$}
 & \multicolumn{2}{c}{Parallax$^{\dagger}$(mas)}
 & \multicolumn{4}{c}{Absolute proper motion$^{\ddagger}$(mas yr$^{-1}$)} \\
 
 &(km s$^{-1}$)
 & \multicolumn{9}{c}{\ \hrulefill \ }                                                                           
 & \multicolumn{2}{c}{\ \hrulefill \ }
 & \multicolumn{4}{c}{\ \hrulefill } \\                                                                     
 
 &
 &1& 2& 3& 4& 5& 6& 7& 8& 9   
 &$\pi$
 &$\sigma _{\pi}$ 
 &$\mu_{X}$ &$\sigma \mu_{X}$ &$\mu_{Y}$ &$\sigma \mu_{Y}$\\                      
\hline
12a\ \dotfill &12.6
 &\ $\circ$&\ $\circ$&\ $\circ$&\ $\circ$&\ $\circ$&\ $\times$&\ $\circ$&\ $\circ$&\ $\times$
 &0.578&0.120&1.01&0.13&$-$2.08&0.16\\
12b\ \dotfill &12.2
 &\ $\circ$&\ $\circ$&\ $\circ$&\ $\circ$&\ $\circ$&\ $\times$&\ $\circ$&\ $\circ$&\ $\times$
 &0.632&0.108&1.02&0.13&$-$2.00&0.16\\
12c\ \dotfill &11.8
 &\ $\circ$&\ $\circ$&\ $\circ$&\ $\circ$&\ $\circ$&\ $\circ$&\ $\circ$&\ $\circ$&\ $\circ$
 &0.641&0.066&1.13&0.10&$-$1.94&0.12\\               
12d \ \dotfill &11.4
 &\ $\circ$&\ $\circ$&\ $\circ$&\ $\circ$&\ $\circ$&\ $\circ$&\ $\circ$&\ $\circ$&\ $\circ$
 &0.667&0.100&1.06&0.10&$-$1.69&0.12\\               
12e\ \dotfill &11.0
 &\ $\circ$&\ $\circ$&\ $\circ$&\ $\circ$&\ $\circ$&\ $\circ$&\ $\circ$&\ $\circ$&\ $\circ$
 &0.678&0.129&1.01&0.10&$-$1.66&0.12\\ 
  \multicolumn{11}{l}{\ \ \ 12 \ Average [Standard deviation]}&\multicolumn{2}{c}{}&1.05&0.26 [0.01]&$-$1.88&0.31 [0.17]\\
\hline
 10a\ \dotfill &10.1
 &\ $\times$&\ $\times$&\ $\circ$&\ $\circ$&\ $\circ$&\ $\times$&\ $\circ$&\ $\times$&\ $\times$
 &1.144&0.546&0.29&0.27&$-$1.00&0.32\\               
10b\ \dotfill &9.7
 &\ $\times$&\ $\circ$&\ $\circ$&\ $\circ$&\ $\circ$&\ $\circ$&\ $\circ$&\ $\circ$&\ $\times$
 &0.683&0.071&0.31&0.15&$-$1.28&0.18\\               
10c\ \dotfill &9.3
 &\ $\times$&\ $\circ$&\ $\circ$&\ $\circ$&\ $\circ$&\ $\times$&\ $\circ$&\ $\circ$&\ $\times$
 &0.774&0.142&0.36&0.15&$-$1.41&0.18\\ 
   \multicolumn{11}{l}{\ \ \ 10 \ Average [Standard deviation]}&\multicolumn{2}{c}{}&0.32&0.34 [0.03]&$-$1.23&0.57 [0.17]\\
\hline
 3a\ \ \dotfill &$-$1.7
 &\ $\circ$&\ $\times$&\ $\circ$&\ $\circ$&\ $\times$&\ $\times$&\ $\circ$&\ $\circ$&\ $\circ$
 &0.617&0.095&1.05&0.11&$-$2.65&0.13\\               
 \hline
 \multicolumn{9}{l}{\ \ \ \ \ \ \ \ Combined fit}&\multicolumn{2}{c}{}&0.659&0.036&&&&\\
\hline
\end{tabular}
\end{center}
\hspace{2.2cm}\parbox[l]{13.7cm}{\footnotemark[${\ast}$]\hspace{0.5mm}\scriptsize ID of maser spot. A Greek letter ID is assigned to each spot in the feature designated in the feature ID. See the maser feature designation in Table 3.}\\
\hspace{2.2cm}\parbox[l]{13.7cm}{\footnotemark[${\dagger}$]\hspace{0.5mm}\scriptsize Detected annual parallax.}\\
\hspace{2.2cm}\parbox[l]{13.7cm}{\footnotemark[${\ddagger}$]\hspace{0.5mm}\scriptsize Spot motion on the plane of sky in the R.A. and Decl directions, respectively. 
The results here were obtained by fitting with assuming an annual parallax of 0.66 mas.}\\
\end{table*}

\subsection{Annual parallax of  \g192}
\label{sec:astrometry}

To measure the annual parallax of G192 we focused our attention on the motions of nine maser spots with respect to the position 
reference J0603. These maser spots are contained within the reference feature G192:S2011-{\it12} as well as in the features 
G192:S2011{\it-10} and {\it-3} (see Table 4); they were successively detected during a period longer than a year, so they were suitable for 
the annual parallax measurement (see Figure \ref{fig:2}). Table \ref{tab:astrometry} shows the measured parallaxes and absolute 
proper motions of the individual spots. The adopted position errors were derived from the position uncertainties that rescaled so as to set the reduced $\chi^{2}$ value to unity in the least-squares fitting. This is necessary because the position errors that were originally 
determined by the Gaussian fits to the brightness distributions underestimate the true position uncertainty. The causes of the position uncertainty 
have been discussed in previous papers based on VERA astrometry. In the present work for a relatively near object, they are mainly attributed to the temporal variation of the 
brightness structure in the maser spots. The final position error resulted to be 0.15 mas in right ascension and 0.19 mas in 
declination. Figure \ref{fig:2} shows the best-fit model to the motion of each spot. Each color corresponds to a different maser spot 
(for which also a letter is assigned, see Table~\ref{tab:astrometry}). The label number indicates the ordinal number of 
the maser feature to which the spot belongs. In addition, we performed a combined fit, in which a common annual parallax 
was estimated from all of the maser spot motions with different linear motion vectors. We adopted the result of the combined fit 
as the final one for the annual parallax determination, which gives a value $\pi = 0.66\pm 0.04$~mas, corresponding to a 
distance to G192 of $D = 1.52 \pm 0.08$~kpc.

\subsection{Motion of  \g192 in the Galaxy}

We estimated the location and the three-dimensional motion of G192 in the Galaxy, based on its derived distance, proper 
motion, and radial velocity. The absolute systemic motion of G192 is derived from the the average absolute proper motion of 
the reference feature plus the relative systemic motion of G192 with respect to the reference feature  ($\bar{\mu}_{x}$, $\bar{\mu}_{y}$)=($-$$0.36\pm 0.14, 0.30\pm 0.14$) [mas~yr$^{-1}$]. 
The former motion is calculated as the average of the five maser spot motions in the reference feature, yielding a motion of 
(${\mu}_{\rm{ref}, x}$,~${\mu}_{\rm{ref},y}$) = (1.05$\pm$0.05, $-$1.87$\pm$0.06) [mas~yr$^{-1}$]. Thus, we obtained the G192 
motion of (${\mu}_{\rm{G192},x}$, ${\mu}_{\rm{G192},y}$) = (0.69$\pm$0.15, $-$1.57$\pm$0.15) [mas~yr$^{-1}$]. Here, we 
adopted the error values derived from the dispersion of the proper motions of the maser spots in reference feature as the 
uncertainty of the G192 motion. The mean Doppler velocity of  about \vlsr~=~5.7\kms\ is consistent within 1~km~s$^{-1}$ with
 that derived from the CO molecular outflow seen on a much larger scale (e.g., SK99). Therefore, we adopt \vlsr = 5.7\kms\ as the 
 systemic velocity. The derived systemic motion vector is converted to $(\mu_{l}\cos(b),\mu_{b})=$ (1.73,$-$0.16) [mas~yr$^{-1}$] 
 in galactic coordinates.

To find the systemic motion vector in the Galaxy, we adopted the solar motion vector, ($U_{\odot}$, $V_{\odot}$, $W_{\odot}$)=(10.0 $\pm$ 0.4, 5.2 $\pm$ 0.6, 7.2 $\pm$ 0.4) [km s$^{-1}$] based on the {\it HIPPARCOS} satellite 
data \citep{deh98}. Here the $U, V, W$ axes point towards the galactic center, the direction of galactic rotation, and the northern 
galactic pole, respectively. We also adopted the IAU standard values for the galactic constants,  $R_{0}$ = 8.5 kpc (the distance 
to the Sun from the Galactic center) and ${\it \Theta}_{0}=$ 220\kms\ (the circular rotation speed of the LSR). 

With the distance of $D$ = 1.52$\pm$0.08 kpc and the radial velocity of \vlsr $=$ 5.7\kms\ obtained in the present paper, we estimate the 
peculiar motion of G192 relative to a circular Galactic rotation as follows, 
\begin{eqnarray*}
&R_{\ast}& = 9.99 \pm 0.08\: \rm kpc, \\
&Z_{\ast}& = -0.10 \pm 0.01\: \rm kpc, \\
&U_{\ast}& = -2.8 \pm 1.0\: \rm km\ s^{-1}, \\
&V_{\ast}& = -10.5 \pm 0.3\: \rm km\ s^{-1}, \\
&W_{\ast}& = 4.9 \pm 2.7\: \rm km\ s^{-1}, 
\end{eqnarray*}
\noindent
where $R_{\ast}$ is the distance to G192 from the galactic center and $Z_{\ast}$ the distance to G192 from the galactic mid-plane. 
The uncertainty of peculiar motion is derived from those of the solar motion and the systemic motion of G192. 
These results suggest that G192 is located at the Perseus spiral arm, which is spread in the range of Galactocentric distance 
of 10--11 kpc in this direction \citep{rei09,oh10}. We also note that G192 is located 100 pc below the galactic mid-plane, which may give 
a lower limit to the thickness of the ``thin disk" component as a site of recent star formation. The origin of the peculiar motion is 
discussed in the next section. 

\section{Discussion}
\subsection{Re-estimating the southern YSO's mass}

In this section, we adopt a value of 1.52 kpc as the distance to G192, which was determined from the annual parallax in substitution of the previous uncertain 
kinematic distance ($\sim$2~kpc; \cite{she96}) to re-estimate the physical parameters of the southern YSO. Firstly, 
we re-scaled the stellar luminosity of G192 from $L_{\ast}\sim3\times 10^{3} L_{\odot}$ (SWSC98) to $L_{\ast}\sim2\times 10^{3} L_{\odot}$. 
Therefore, the spectral type of G192 should be revised from B2 to spectral type B3. Secondly, we re-estimated the mass of the disk plus 
circumstellar envelope of the southern YSO, which was originally estimated by SK99 from observations of millimeter continuum emission. Using equation (1) of  SK99 (see also \cite{hil83}),  these authors mention that this value should be constrained within the range $7\:M_{\odot}\leq M_{\mbox{\scriptsize gas+dust}} \leq 32\:M_{\odot}$ (c.f. SWSC98). 
Note that the estimated mass of the southern YSO is proportional to square of the distance, in this equation.
By adopting the distance obtained from our work, we modified this value to be $4\:M_{\odot} \leq M_{\mbox{\scriptsize gas$+$dust}} \leq 18\:M_{\odot}$. 
We have also modified the mass value of the 7 mm continuum emission source from 6.7$\pm 1.0\:M_{\odot}$ to 3.9$^{+1.0}_{-0.9}\:M_{\odot}$.

This suggests that the southern YSO is an intermediate-mass star but it is still consistent with the existence of a giant outflow and Herbig-Haro objects, as 
those actually observed, driven by such a YSO. On the other hand, it should be re-evaluated whether the centimeter emission originates 
from an H{\rm II} region as seen in MYSOs. Thus, although it is difficult to explore G192 as a site of massive star formation, it is still an 
interesting source, which may directly exhibit the mass accretion onto a YSO that may be traced by \h2o maser emission (Paper I). 

\subsection{The origin of the peculiar motion of \g192}

The slower galactic rotation velocity of G192 by $\sim$10\kms\ with respect to the local standard of rest in the Perseus Arm is similar to the 
cases of S252 \citep{rei09} and IRAS 06058+2138 \citep{oh10}, which are located at a similar Galactic longitude and in the Perseus arm. 
The new distance value for G192 is consistent with the existence of a "dip profile" in the Galactic rotation curve as shown by previous 
observations (\cite{fic89}; \cite{hon97}; \cite{dem07}; \cite{sof09}; \cite{oh10}). The dip profile is seen from 8.5 to 11~kpc from the observations 
of the HI and CO molecular emission. The outer part of the rotation curve exhibits a flat profile. 
It is noteworthy to mention that G192 seems to be located in the "down stream" of the density wave in the spiral arm, where newly-born stars are 
concentrated. However, such discussion is still based on a small sample of sources in a small portion of the Perseus arm. Therefore, 
more observations to several directions are necessary to examine the existence of a dip profile. Note that the peculiar motion of G192 with 
respect to the Galactic rotation still remains within those widely seen in other \h2o\ maser sources. We note that the northern YSO is 
relatively approaching the southern YSO with a velocity of $\sim$12\kms, assuming that the latter follows a circular galactic rotation. We 
pondered whether this motion is explained by an orbital motion. First, we supposed the case in which the northern YSO and the southern 
one are gravitationally bound. Since the northern YSO in G192 has a lower mass than the southern YSO (SK99), we suppose that the 
former should be orbiting the latter according to Kepler's law. Although the real separation between the YSOs is unknown, a minimum value
of $\sim$900 AU is adopted (see Figure 2 of Paper I). Thus a lower limit to the enclosed mass is calculated to be 150~$M_{\odot}$, which 
turns out to be much higher than that estimated in the previous section. Therefore, we conclude that the observed peculiar motion may 
be attributed to the Galactic dynamics rather than to the orbiting motion of the northern YSO around the southern one. However, the motion 
of the southern YSO should be directly measured to obtain an unambiguous conclusion. 

\subsection{The outflow development in the northern YSO}
SBCK04 and Paper I suggest that there are two parallel outflows in G192, one of which is highly collimated and almost completely parallel 
to the maser jet. It is also expected that the northern and southern objects started their star-formation process from a common molecular 
cloud core and these outflows simultaneously started so as to have the same dynamical age. Thus, the evolution speed is surely 
different, but the elapsed time after the star formation starts in northern and southern objects is the same. Therefore, we can regard that the 
physical parameters of the outflows are similar even in the case that the mass of the objects are different. In addition, we adopt the extent of HH objects as the 
true size of the flow rather than the size of the distribution of the maser spot. Based on the 3D kinematical structure of the \h2o masers that 
are associated with the outflow driven by the northern YSO, we estimated the dynamical ages of the bipolar outflows in G192. We estimated 
an upper limit of the lengths of these outflows projected on the sky to $\sim$5.7~pc from the existence of the Herbig-Haro objects HH 396/397 
(\cite{dev99}, see also the $^{12}$CO emission map of SWSC98). We also estimated the velocity of the outflows of $\sim$40\kms\ from the 
motion of the blue-shifted maser feature  G192:S2011-{\it11}, which is the one moving at the highest speed in the vicinity of the flow's kinematical 
center (see figure \ref{fig:1}). Assuming a constant velocity in the whole outflow, the dynamical age of the outflows is estimated to be 
$\sim 1.4\times 10^{5}$~yr. This value is shorter by a factor of 1.5 than that previously estimated value without any information of the 3-D 
flow velocity ($\sim$2.0$\times$10$^{5}$~yr, \cite{sne90}; SWSC98; \cite{dev99}). This suggests that G192 is either in the late stage of 
massive star formation (in the case that a MYSOs really exist) or in the phase of main mass accretion onto a lower mass YSO with the 
development of highly collimated outflows. We note that the derived time scale is comparable to the whole duration of \h2o\ maser activity 
\citep{gen77} although it looks much longer than those of the outflows traced by \h2o\ maser emission in other massive-star forming 
regions ($<10^{4}$~yr). The coexistence of \h2o\ maser features in both the northern and southern YSOs with different stellar mass and 
evolutionary phases may support the time scale of $\sim 10^5$~yr. 

\section{Conclusions}

We have measured an annual parallax of G192, $\pi = 0.66\pm0.04$~mas, corresponding to a distance $D = 1.52\pm 0.08$~kpc from the Sun, 
using VLBI astrometric observations with VERA in a time span of nearly two years.  We have also derived the angular distribution and three-dimensional 
velocity field of \h2o masers in the northern YSO of G192. From the present observations, we have re-estimated the mass of the disk plus circumstellar 
cloud in the southern cluster and concluded that they may be an intermediate-mass YSO. The \h2o masers trace a bipolar outflow with a major axis 
along the northeast--southwest direction and are located at the northern parts in G192. The outflow found in this region is well explained as to be driven 
by an intermediate-mass YSO as seen in Herbig-Haro objects. The dynamical age of the outflows is also estimated to be $\sim 1.4\times 10^{5}$~yr 
(shorter by a factor of 1.5 than those previously estimated), suggesting that this source may be in the phase of main mass accretion with 
development of highly collimated outflows. In addition, we have obtained the location and the peculiar motion of the object with respect to a circular 
Galactic rotation. The peculiar motion of G192 is likely to be within the typical values as those widely found in other MYSO \h2o\ maser sources. 
However, it should be further explored for understanding the relation between the star formation process and the galactic motion. On the other hand, 
we could not detect internal motions in the southern maser clusters, which should be explored in future VLBI observations when the target maser 
components are visible. 

\bigskip

HI has been financially supported by Grant-in-Aid for Young Scientists from the Ministry of Education, Culture, Sports, Science, and Technology 
(18740109) as well as by Grant-in-Aid for Scientific Research from Japan Society for Promotion Science (20540234). DT acknowledges support 
from the Japan Society for Promotion of Science (project ID: 22-00022)














\begin{thebibliography}{}
\bibitem[Baba et al.(2009)]{bab09}Baba,~J., Asaki,~Y., Makino,~J., Miyoshi,~M., Saitoh,~T.~R., \& Wada,~K., 2009, \apj, 706, 471 Cesaroni,~R., Felli,~M., Jenness,~T., Neri,~R., Olmi,~L., Robberto,~M., Testi,~L., \& Walmsley,~C. M. 1999, A\&A, 345, 949
\bibitem[Cesaroni et~al.(1999)]{cea99}Cesaroni, R., Felli, M., Jenness, T., Neri, R., Olmi, L., Robberto, M., Testi, L., \& Walmsley, C. M. 1999, A\&A, 345, 949
\bibitem[Dehnen \& Binney(1998)]{deh98}Dehnen, W., \& Binney, J. J. 1998, MNRAS, 298, 387
\bibitem[Demers \& Battinelli(2007)]{dem07}Demers, S., \& Battinelli, P. 2007 A\&A, 473, 143
\bibitem[Devine et~al.(1999)]{dev99}Devine,~D., Bally,~J., Reipurth,~B., Shepherd,~D.~S., \&  Watson, ~A.~M.,  1999, \aj, 117, 2919
\bibitem[Diamond(1995)]{dia95}Diamond,~P.~J., 1995, in ASP Conf.~Ser.~82, VERY LONG BASELINE INTERFEROMETRY AND THE VLBA, ed.,  J.~A.~Zensus, P.~J.~Diamond, \&  P.~J.~Napier (San Francisco: ASP), p.~227
\bibitem[Fich et~al.(1989)]{fic89}Fich, M., Blitz, L., \& Stark, A. A. 1989, ApJ, 342, 272
\bibitem[Genzel \&  Downes(1977)]{gen77}Genzel,~R., \&  Downes,~D., 1977, \aaps, 30, 145
\bibitem[G\'omez et~al.(2006)]{gom06}G\'omez,~G.~C. 2006, AJ, 132, 2376
\bibitem[Hildebrand(1983)]{hil83}Hildebrand, R. H. 1983, QJRAS, 24,267
\bibitem[Hirota et~al.(2007)]{hir07}Hirota,~T., et~al.,  2007, \pasj, 59, 897
\bibitem[Honma \& Sofue(1997)]{hon97}Honma, M., \& Sofue, Y. 1997, PASJ, 49, 453
\bibitem[Honma et~al.(2008)]{hon08}Honma,~M., et~al.,  2008, \pasj, 60, 935 
\bibitem[Honma et~al.(2003)]{hon03}Honma,~M., et~al.,   2003, \pasj, 55, L57
\bibitem[Honma et~al.(2007)]{hon07}Honma,~M., et~al.,  2007, \pasj, 59, 889
\bibitem[Imai et~al.(2011)]{ima11}Imai,~H., et~al., 2011, \pasj, 63, in press 
\bibitem[Imai et~al.(2006)]{ima06}Imai,~H., et~al., 2006, \pasj, 58, 883 (Paper I)
\bibitem[Indebetouw et~al.(2003)]{ind03}Indebetouw,~R., Watson,~C., Johnson,~K.~E., Whintney,~B., \&  Churchwell,~E.,  2003, \apj, 596, L83
\bibitem[Moscadelli et~al.(2009)]{mos09}Moscadelli,~L., Reid,~M.~J., Menten,~K.~M., Brunthaler,~A., Zheng,~X.~W., \& Xu,~Y.,  2009, \apj, 693, 406
\bibitem[Nagayama et~al.(2011)]{nag11}Nagayama,~T., et~al.,  2011, \pasj, 63, in press 
\bibitem[Nakagawa et~al.(2011)]{nak11}Nakagawa,~A., et~al.,  2011, \pasj, 63, in press
\bibitem[Nakagawa et~al.(2008)]{nak08}Nakagawa,~A. et~al.,  2008, \pasj, 60, 1013
\bibitem[Oh et~al.(2010)]{oh10}Oh,~C.~-S., Kobayashi,~H., Honma,~M., Hirota,~T., Sato,~K., \& Ueno,~Y.\ 2010, \pasj, 62, 101
\bibitem[Patel et~al.(2005)]{pat05}Patel,~N. A., et~al.,  2005, Nature, 437, 109 
\bibitem[Reid et~al.(2009)]{rei09}Reid,~M.~J., Menten,~K.~M., Brunthaler,~A., Zheng,~X.~W., \& Xu,~Y. 2009, ApJ, 693, 397
\bibitem[Sato et~al.(2010)]{sat10}Sato,~M. et~al.,  2010, \pasj, 62, 287
\bibitem[Shepherd et~al.(2004)]{she04}Shepherd,~D.~S., Borders,~T., Claussen,~M.~J., Shirley,~Y.\ \&  Kurtz,~S.~E.\ 2004, \apj, 614, 211 (SBCSK04)
\bibitem[Shepherd, Claussen, \&  Kurtz(2001)]{she01}Shepherd,~D.~S., Claussen,~M.~J., \& Kurtz,~S.~E.\ 2001, Science, 292, 1513 
\bibitem[Shepherd \&  Kurtz(1999)]{she99}Shepherd,~D.~S., \& Kurtz,~S.~E.,  1999, \apj, 523, 690 (SK99)
\bibitem[Shepherd  et~al.(1998)]{she98}Shepherd,~D.~S., Watson,~A.~M., Sargent,~A.~I., \&  Churchwell,~E.\ 1998, \apj, 507, 861 (SWSC98)
\bibitem[Shepherd \& Churchwell(1996)]{she96}Shepherd,~D.~S.,  \&  Churchwell,~E.,  1996, \apj, 472, 225
\bibitem[Snell et~al.(1990)]{sne90}Snell,~R.~L., Dickman,~R.~L., \&  Huang,Y.-L.,  1990, \apj, 352, 139
\bibitem[Sofue et~al.(2009)]{sof09}Sofue, Y., Honma, M., \& Omodaka, T. 2009, PASJ, 61, 227
\bibitem[Thompson(1984)]{tho84}Thompson, R. I. 1984, ApJ, 283, 165
\bibitem[Xu et~al.(2006)]{xu06}Xu,~Y., Reid,~M.~J., Zheng,~W.~W., \& Menten,~K.~M., 2006, Science, 311, 54
\bibitem[Zhang et~al.(2002)]{zha02}Zhang,~Q., Hunter,~T.~R., Sridharan,~T.~K., \& Ho,~P.~T.~P., 2002, ApJ, 566, 982
\end{thebibliography}
\end{document}